\documentclass[12pt]{iopart}

\usepackage{graphicx}

\usepackage{xspace}

\newcommand{\degree}{\ensuremath{{}^{\circ}}C\xspace}
\newcommand{\sups}[1]{$^{#1}$} %superscript
\newcommand{\subs}[1]{$_{#1}$} %subscript
\newcommand{\RECCO}{{\it RE}\subs{2-x}Ce\subs{x}CuO\subs{4}\xspace}
\newcommand{\NCCO}{Nd\subs{2-x}Ce\subs{x}CuO\subs{4}\xspace}
\newcommand{\PCCO}{Pr\subs{2-x}Ce\subs{x}CuO\subs{4}\xspace}
\newcommand{\LSCO}{La\subs{2-x}Sr\subs{x}CuO\subs{4}\xspace}
\newcommand{\YBCO}{YBa\subs{2}Cu\subs{3}O\subs{7-\delta}\xspace}
\newcommand{\LSAT}{[LaAlO\subs{3}]\subs{0.3}[Sr\subs{2}AlTaO\subs{6}]\subs{0.7}\xspace}
\newcommand{\CNO}{(Nd,Ce)\subs{2}O\subs{3}\xspace}
\newcommand{\CRO}{({\it RE},Ce)\subs{2}O\subs{3}\xspace}
\newcommand{\NCCOopt}{Nd\subs{1.85}Ce\subs{0.15}CuO\subs{4}\xspace}
\newcommand{\NCCOoptp}{Nd\subs{1.85}Ce\subs{0.15}Cu\subs{1.1}O\subs{4}\xspace}

\begin{document}
\title[Effect of high oxygen pressure annealing on superconducting \NCCOopt\dots]{Effect of high oxygen pressure annealing on superconducting \NCCOopt thin films by pulsed laser deposition from Cu-enriched targets}
\author{M Hoek, F Coneri, D P Leusink, P D Eerkes, X Renshaw Wang and H Hilgenkamp$^{\footnotemark[1]}$}
\footnotetext{also at: Institute-Lorentz, Leiden University, 2300 RA Leiden, The Netherlands}
\address{ Faculty of Science and Technology and MESA+ Institute for Nanotechnology, University of Twente, 7500 AE Enschede, The Netherlands}
\ead{m.hoek@utwente.nl}

\date{\today}

\begin{abstract}
We show that the quality of \NCCOopt films grown by pulsed laser deposition can be enhanced by using a non-stoichiometric target with extra copper added to suppress the formation of a parasitic \CNO phase. The properties of these films are less dependent on the exact annealing procedure after deposition as compared to films grown from a stoichiometric target. Film growth can be followed by a 1 bar oxygen annealing, after an initial vacuum annealing, while retaining the superconducting properties and quality.  This enables the integration of electron-doped cuprates with their hole-doped counterparts on a single chip, to create, for example, superconducting {\em pn}-junctions.

\end{abstract}
%\submitto{\SUST}
\maketitle

\section{Introduction}
With the substitutional doping in cuprates by Ce in electron-doped \NCCO and Sr in hole-doped \LSCO, the parallel to semiconductors is easy to make. Indeed, various theoretical analyses have been made on the properties and possibilities of combinations of electron- and hole-doped cuprates, ranging from the formation of a Mott insulator depletion zone \cite{Charlebois2013}, unconventional Josephson junctions \cite{Mannhart1993,Hu2007} and superradiant light emission \cite{Hanamura2002}. Oxygen plays an important role in the realization of combinations of electron- and hole-doped cuprates. The role of oxygen in {\em n-}type cuprates is widely researched and debated \cite{Schultz1996,Matsuura2003,Richard2004,Riou2004,Higgins2006,Gauthier2007,Kang2007,Armitage2010}. The consensus is that an oxygen reduction is necessary for superconductivity and that strong oxygenation leads to loss of superconductivity and increasing resistance. However, the exact mechanism for oxygen reduction is still under debate, in particular where oxygen is removed from the unit cell and what the consequences are for the structure of the cuprate \cite{Higgins2006,Gauthier2007,Kang2007,Roberge2009}. The necessity of a reduction is in stark contrast with the oxygenation needed for optimal superconductivity in the {\em p-}type cuprates: in general, {\em p-}type cuprates require an oxygenation step at high oxygen pressures. These conflicting annealing requirements are one of the main hurdles to realize combinations of electron- and hole-doped cuprates.   

In this work we investigate the effect of strong oxygenation, needed for the growth of a {\em p-}type cuprate like \LSCO, on the {\em n-}type cuprate \NCCO. Growth of \RECCO ({\it RE} = Nd, Pr, \ldots ) as single crystals grown by e.g. the traveling-solvent floating-zone technique \cite{Kang2007,Kurahashi2002,Mang2004,Kimura2005} or as thin films by pulsed laser deposition (PLD)   \cite{Armitage2010,Roberge2009,Beesabathina1993,Mao1994} is almost always accompanied by the formation of a parasitic phase of \CRO, due to copper deficiency during growth and during oxygen reduction. Other techniques like molecular beam epitaxy \cite{Naito2002} and dc sputtering \cite{Guarino2012} appear to be less sensitive to the formation of the \CRO phase. For PLD, it has been shown that this parasitic phase can be suppressed in \PCCO by adding extra copper to the PLD target \cite{Roberge2009}. We show that the parasitic \CNO phase in \NCCOopt can also be suppressed by using copper-rich targets and that these films retain their quality and superconducting properties when subjected to oxygen annealing procedures suitable for the growth of {\em p-}type cuprates.

\section{Experimental details}
We compare films grown by PLD using two different targets with effective compositions \NCCOopt (NCCO) and \NCCOoptp (NCCO+), the latter containing 10\% extra copper. The targets are prepared by solid state synthesis and the crystal structure is verified by X-ray powder diffraction. The films are deposited at a heater temperature of 820~\degree in 0.25 mbar oxygen using a KrF excimer laser with a fluency of 1.2 J/cm\sups{2}, a spot size of 5.7 mm\sups{2} and a repetition rate of 4 Hz. All films are grown on (001) oriented \LSAT (LSAT) substrates that have been annealed in flowing oxygen for 10 hours at 1050~\degree to obtain an atomically smooth surface, as measured by atomic force microscopy (AFM). We have grown films varying in thickness between 30 and 500 nm, with most of the films being 70 nm. We have chosen this value with future device fabrication requirements in mind, although with this thickness the lattice strain from the substrate slightly suppresses $T_c$ \cite{Mao1994}.

We have investigated three different annealing procedures for the \NCCOopt films, see table \ref{tab:ann}. The first is a standard vacuum annealing where the film is cooled down from 820~\degree in deposition pressure and then annealed in vacuum for 8 minutes at 740~\degree followed by a cool down in vacuum at 10~\degree/min (procedure {\em standard} in table \ref{tab:ann}). The other two procedures are two opposite cases, to explore the stability of the \NCCOopt films under long reduction and strong oxygenation, respectively (procedures {\em long} and {\em oxygen} in table \ref{tab:ann}).  For the long reduction, the dwell time at 740~\degree is extended to 45 minutes. For the strong oxygenation procedure there is an initial 8 minute vacuum annealing at 740~\degree and then at 600~\degree the conditions are changed to a common recipe we employ for {\em p-}doped superconductors like \LSCO and \YBCO. Here, the film is first annealed in 1 bar oxygen for 15 minutes at 600~\degree and then for 30 minutes at 450~\degree and subsequently cooled down further in 1 bar oxygen.

The film thickness and quality is investigated by X-ray diffraction (XRD), cross-sectional transmission electron microscopy (TEM) and by AFM.  The thickness is measured by AFM using an edge fabricated by a hard mask lift-off of amorphous \YBCO (lift-off using 1\% H\subs{3}PO\subs{4}) or gold (lift-off using a KI solution). XRD shows that all films are c-axis oriented with a c-axis length of 12.08(1) \AA, this is also what is observed for optimally doped single crystal \cite{Kimura2005}. We measure an RMS surface rougness with AFM on a scan area of 2 {$\mu$m}\,$\times$\,2 {$\mu$m} of 3 nm for the \NCCOopt films grown using the target with extra copper and 1.5 nm for the \NCCOopt films from the stoichiometric target, both for films of 70 nm thickness. 

The samples are contacted at the corners by Al bond wires on Au/Ti contact pads in a standard Van der Pauw geometry for sheet resistance and Hall measurements in a Quantum Design 9 T PPMS system. Hall measurements are performed up to 4 T and for low temperatures up to 9 T. The sheet resistance is measured both during warm up and cool down at a rate varying between 0.1~\degree and 3~\degree per minute.

\begin{table}[ht]
\caption{\label{tab:ann}Overview of the three different annealing procedures, {\em standard}, {\em long} and {\em oxygen}. All procedures start with a cool down in deposition pressure from 820 \degree to 740 \degree and all temperature changes are performed at 10 \degree/min. {\em Cool down II} is performed in vacuum and {\em Oxygen annealing} and {\em Cool down III} in 1 bar oxygen.} 

%\begin{indented}
\lineup
%\item[]
\begin{tabular}{@{}lllll}
\br                              
Procedure & Vacuum annealing & Cool down II & Oxygen annealing & Cool down III\cr 
\mr
{\em standard}  & \08 min. at 740 \degree & 740 \degree to RT$^{\footnotemark[1]}$ & -- & --\cr
{\em long}  & 45 min. at 740 \degree & 740 \degree to RT & -- & --\cr
{\em oxygen} & \08 min. at 740 \degree & 740 -- 600 \degree & 15 min. at 600 \degree, & 450 \degree to RT\cr
{}&{}&{}&30 min. at 450 \degree &{}\cr
\br
\end{tabular}
%\item[] 
$^{\footnotemark[1]}$Room temperature
%\end{indented}

\end{table}

\begin{figure}[ht]

\centering
\includegraphics[width=0.5\textwidth]{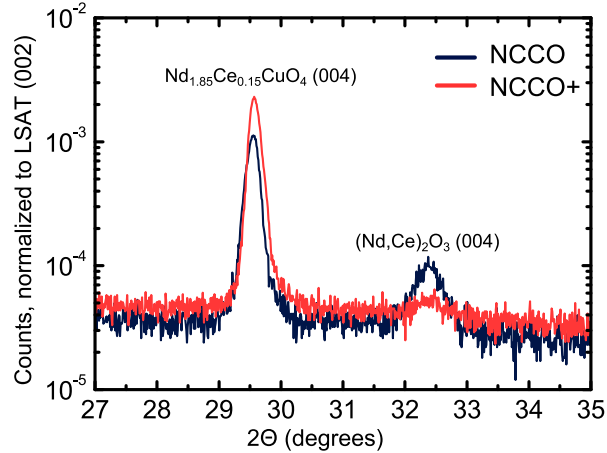}
\caption{Detail of X-ray diffraction $\theta - 2\theta$ scans normalized to the LSAT (002) peak (not shown in figure) on 70 nm \NCCOopt films deposited from the stoichiometric target (NCCO, blue) and the non-stoichiometric target with extra copper added (NCCO+,  red),  showing the suppression of the parasitic \CNO phase. Both samples have been annealed for 8 minutes in vacuum at 740~\degree ({\em standard} procedure).
}
\label{fig:1}

\end{figure}

\section{Results and discussion}
\subsection{Suppression of the \CNO parasitic phase}
\begin{figure}[ht]

\centering
\includegraphics[width=0.4\textwidth]{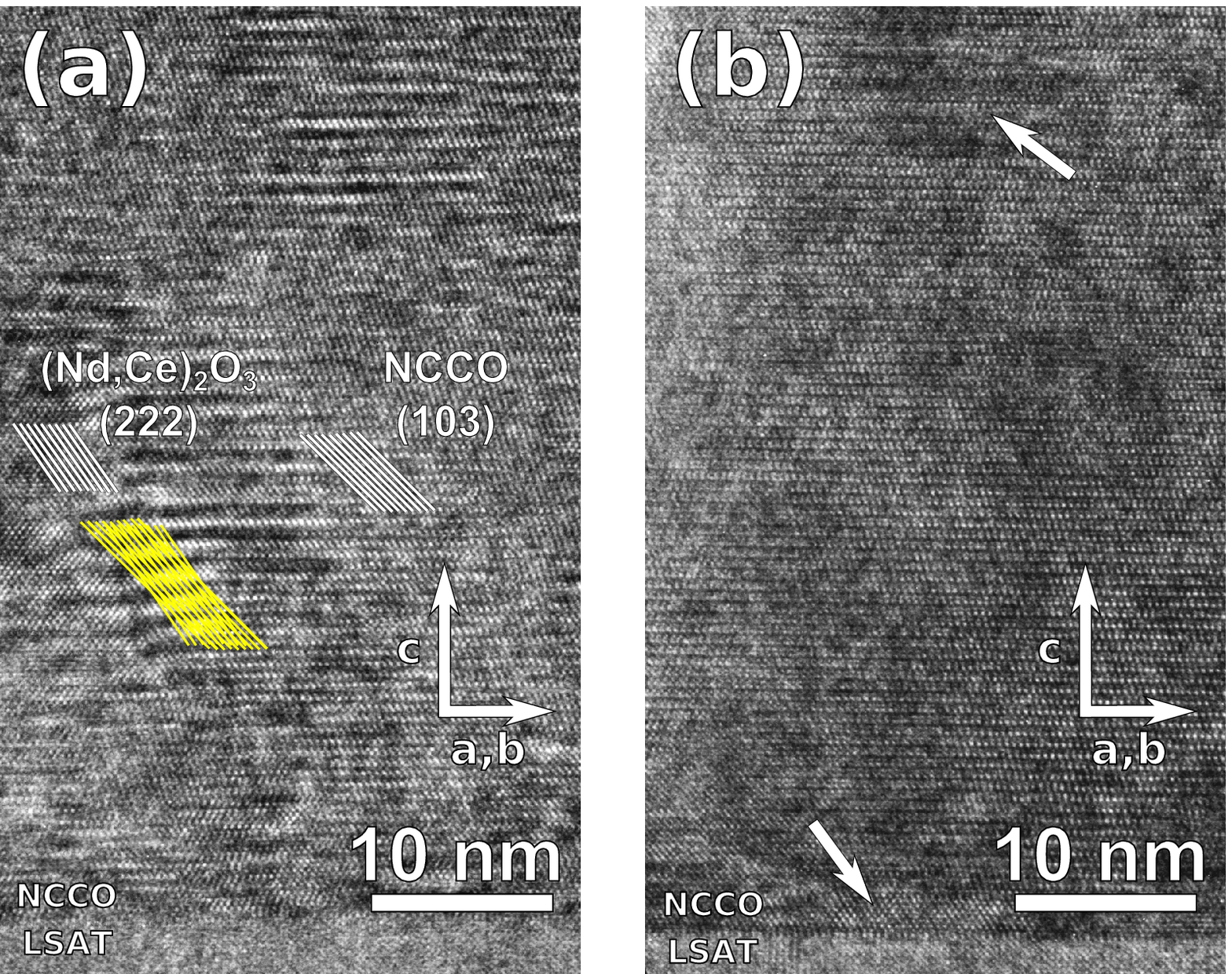}
\caption{Transmission electron microscopy close ups near the substrate of 70 nm \NCCOopt films deposited from the stoichiometric target (a) and the non-stoichiometric target with extra copper added (b), both films are annealed following the {\em standard} procedure, see table \ref{tab:ann}. (a) \NCCOopt from the stoichiometric target shows Moir\'e fringes (indicated by the yellow lines), caused by the overlapping of the  \NCCOopt (103) lattice planes and the \CNO (222) lattice planes, both identified by the white lines; (b) \NCCOopt from the non-stoichiometric target with extra copper shows higher \NCCOopt phase purity with only a small inclusion of the parasitic phase, indicated by the white arrows in the top and the bottom of the figure. 
}
\label{fig:2}

\end{figure}

\begin{figure}[ht]

\centering
\includegraphics[width=1\textwidth]{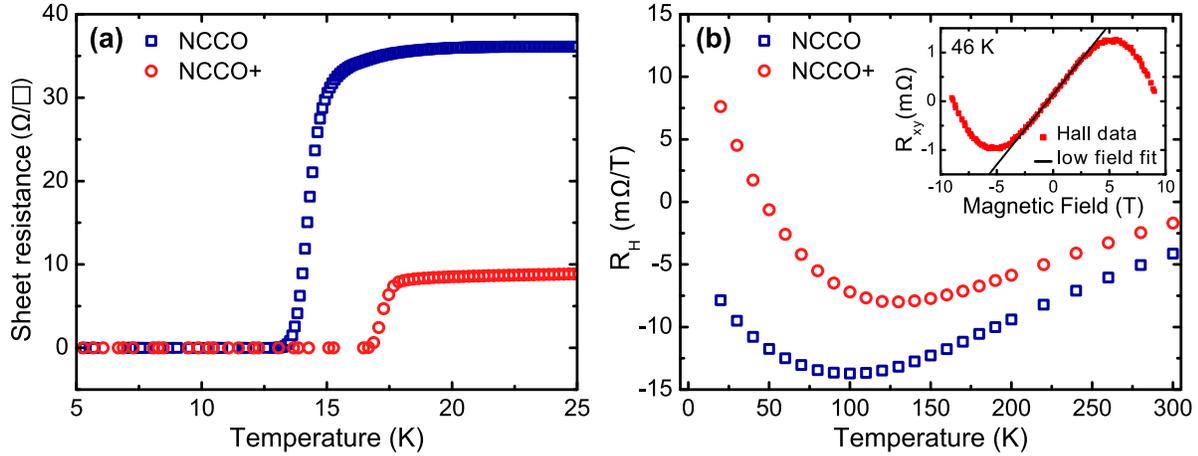}
\caption{Transport measurements on 70 nm \NCCOopt films deposited from the stoichiometric target (NCCO, blue squares) and the non-stoichiometric target with extra copper added (NCCO+, red circles), both films are annealed in vacuum for 8 minutes at 740~\degree ({\em standard} procedure). (a) Sheet resistance around $T_c$, showing a higher $T_c$, sharper transition and lower resistance for the \NCCOopt film with extra copper; (b) Hall coefficient versus temperature measured in the Van der Pauw geometry, both films show a pronounced minimum and the \NCCOopt grown from the target with extra copper shows a crossover to a positive Hall coefficient for low temperature. The inset shows the two-band nature of the Hall resistance at 46 K and a low field linear fit.}
\label{fig:3}

\end{figure}

This section compares films grown using the targets with and without extra copper added. Figures \ref{fig:1}, \ref{fig:2} and \ref{fig:3} compare the same two samples, representative for the general behavior we have observed, both are 70 nm thick and are annealed in vacuum for 8 minutes at 740~\degree after deposition.  Figure \ref{fig:1} shows a detail of the XRD $\theta-2\theta$ spectra for the two \NCCOopt films, grown with and without extra copper in the target, showing the \NCCOopt (004) and the \CNO (004) reflection. The spectra are normalized to the (002) Bragg reflection of LSAT (not shown in the figure).  With the addition of extra copper to the target, we see a reduction of nearly an order of magnitude in the ratio between the intensities of the \CNO (004) diffraction peak and the \NCCOopt (004) peak. Together with this reduction, we observe a higher intensity for the \NCCOopt (004) peak, while the full width at half maximum is slightly decreased (from 0.23${}^{\circ}$ to 0.19${}^{\circ}$), indicating a higher crystallinity for the films grown with extra copper. The \CNO phase is still not fully suppressed, suggesting that an even higher percentage of copper may be required for full suppression. 
We find no appreciable difference in the c-axis lattice parameter between the different annealing procedures for films grown with either of the two targets.

Figure \ref{fig:2} shows TEM close ups of two \NCCOopt films near the substrate/film interface, grown using the two targets without, figure \ref{fig:2}(a), and with extra copper, figure \ref{fig:2}(b), both annealed in vacuum for 8 minutes at 740~\degree. For the film grown using the stoichiometric target, the presence of the parasitic \CNO phase is confirmed by the appearance of Moir\'e patterns in the TEM image. The image shows the crystal planes of \NCCOopt and areas with a different periodicity; here, a Moir\'e pattern is formed by an overlap of the \NCCOopt lattice with the lattice of the \CNO parasitic phase, namely the \NCCOopt (103) lattice planes and the \CNO (222) lattice planes. The (103) and the (222) planes are sketched in figure \ref{fig:2}(a), as well as a set of overlapping planes on top of one of the Moir\'e patterns. The planes are not arbitrary, \NCCOopt (103) and \CNO (222) are also the planes that would give the strongest signal in X-ray powder diffraction. The periodicity of the Moir\'e pattern ($d_M$) can be calculated from the difference between the g-vectors of the individual lattices \cite{Williams1996}: 

\begin{equation}
d_M = \frac{d_{(103)}d_{(222)}}{\sqrt{d_{(103)}^2+d_{(222)}^2-2d_{(103)}d_{(222)}\cos{\beta}}},
\end{equation}
where $d_{(103)}$ and $d_{(222)}$ are the lattice spacing of the \NCCOopt (103) planes and the \CNO (222) planes, respectively, and $\beta$ is the angle in radians between the g-vectors, normal to these planes. Using the lattice parameters for \NCCOopt and \CNO from Kimura {\it et al.} \cite{Kimura2005} ($d_{(103)}=2.82$ \AA, $d_{(222)}=3.22$ \AA{} and $\beta = 0.16$ rad), we find a Moir\'e pattern spacing of 14.4 \AA, which is also what we observe in the TEM images.

With the addition of extra copper to the target, we observe a significantly lower density of the parasitic phase in the film. Figure \ref{fig:2}(b) is a typical example, showing a higher \NCCOopt phase purity with only a small inclusion of the parasitic phase at the substrate interface and in the top of the image (indicated with white arrows). As was also observed in XRD, see figure \ref{fig:1}, the parasitic phase is strongly, but not completely, suppressed.

Figure \ref{fig:3} compares transport properties of the \NCCOopt films. The suppression of the parasitic \CNO phase is accompanied by in an increase of $T_c$ by 3.5 K to a value of 16.7~K ($T_{c0}$) for films grown using the non-stoichiometric target, see figure \ref{fig:3}(a). These films also show a sharper transition to the superconducting phase and a lower normal-state resistance. The sharper transition can be explained by a more homogeneous crystal structure for the films grown from the target with extra copper. The higher resistance for the \NCCOopt films grown from the stoichiometric target may be explained by the \CNO phase deforming nearby CuO\subs{2} planes and acting as nucleation site for defects and dislocations, all increasing scattering.

The Hall coefficient for both films shows a trend generally observed for {\em n-}type cuprates around optimal doping, with an upturn towards positive values with decreasing temperature, indicating contribution of hole-like carriers \cite{Armitage2010,Kubo1991,Jiang1994,Charpentier2010}, see figure \ref{fig:3}(b). For \NCCOopt films grown from the target with extra copper, the minimum in the Hall coefficient shifts to a higher temperature. For low temperatures, a complete cross-over to a positive Hall coefficient is observed with a cross-over region where the Hall resistance displays both electron- and hole-like character, as shown in the inset to figure \ref{fig:3}(b). The data in figure \ref{fig:3}(b) only uses a low field linear fit to the Hall data, illustrated by the black linear fit in the inset. The difference between the two curves can be explained by the two band nature of \NCCO around optimal doping. Here the Hall coefficient not only measures the carrier density, but a combination of the density and mobility of both carrier types. As we argue that the suppression of the \CNO phase decreases scattering, this will lead to a higher mobility, which will in turn be reflected in the Hall coefficient.

\begin{figure}[ht]

\centering
\includegraphics[width=1\textwidth]{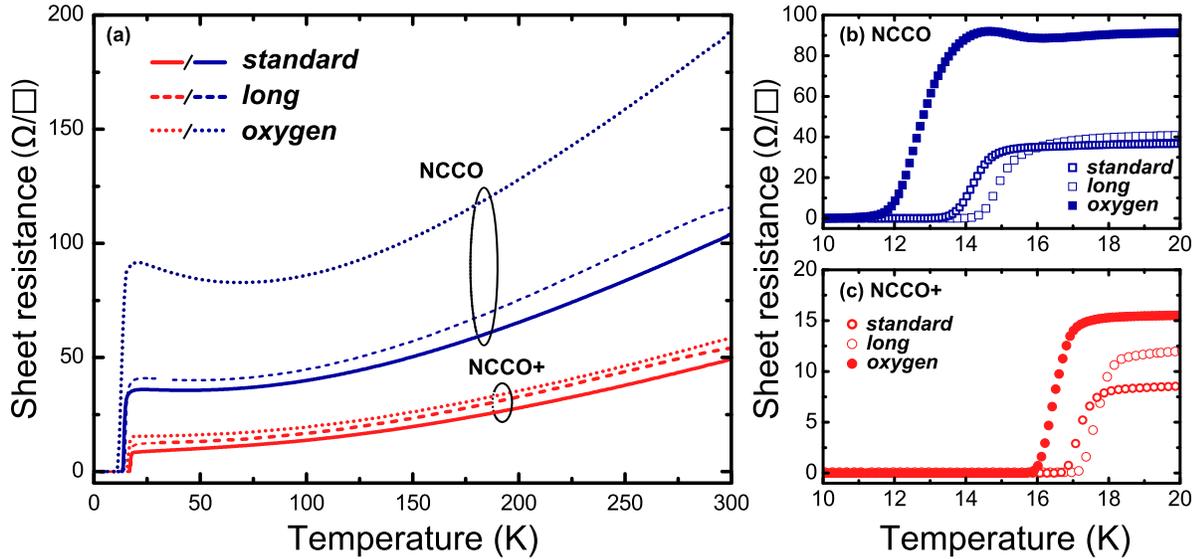}

\caption{(a) Transport measurements on 70 nm \NCCOopt films deposited from the stoichiometric target (NCCO, blue) and the non-stoichiometric target with extra copper added (NCCO+, red) for three different annealing procedures as described in table \ref{tab:ann}: {\em standard} (solid lines), {\em long} (dashed lines) and {\em oxygen}  (dotted lines); (b) closeup around $T_c$ for the \NCCOopt films from the stoichiometric target; (c) closeup around $T_c$ for the \NCCOopt films with extra copper added to the target.}
\label{fig:4}

\end{figure}

\subsection{Effect of different annealing procedures}
Finally, we look at the effect of different annealing procedures for films grown using the stoichiometric target and the target with extra copper added. The three annealing procedures are as described earlier: 8 minutes vacuum ({\em standard}), 45 minutes vacuum ({\em long}), and 8 minutes vacuum followed by 1 bar oxygen at 600~\degree (15 min.) and 450~\degree (30 min.) ({\em oxygen}), see also table \ref{tab:ann}. The sheet resistance for all films across the whole temperature range (2-300 K) is shown in figure \ref{fig:4}(a). The films grown from the target with extra copper show a close grouping of the curves, whereas the films grown with the stoichiometric target show a large spread. The largest deviation is found for the \NCCOopt films grown using the stoichiometric target and annealed in oxygen after the initial vacuum annealing ({\em oxygen} procedure). Here, an upturn in the resistance is observed above $T_c$, indicating a shift to lower doping by oxygen inclusion \cite{Higgins2006}, increased impurity scattering and carrier localization \cite{Jiang1994,Xu1996}.  These samples show a broad superconducting transition with a small upturn at $T_c$, characteristic for sheet measurements on inhomogeneous superconductors with an out-of-line contact arrangement \cite{Vaglio1993}, see figure \ref{fig:4}(b). We find the highest $T_c$ for films annealed for 45 minutes in vacuum.

The films grown with the target with extra copper show only a 2 K spread in $T_c$ with the different annealing procedures and the width of the superconducting transition is always smaller than 1 K. The same trend as for the films without extra copper is observed, with a 45 minute vacuum annealing ({\em long} procedure) giving the highest $T_c$ and a 1 bar oxygen annealing at 600~\degree ({\em oxygen} procedure) the lowest $T_c$, see figure \ref{fig:4}(c). All films with extra copper show a higher $T_c$ than the films grown with the stoichiometric target.

\begin{figure}[ht]

\centering
\includegraphics[width=0.7\textwidth]{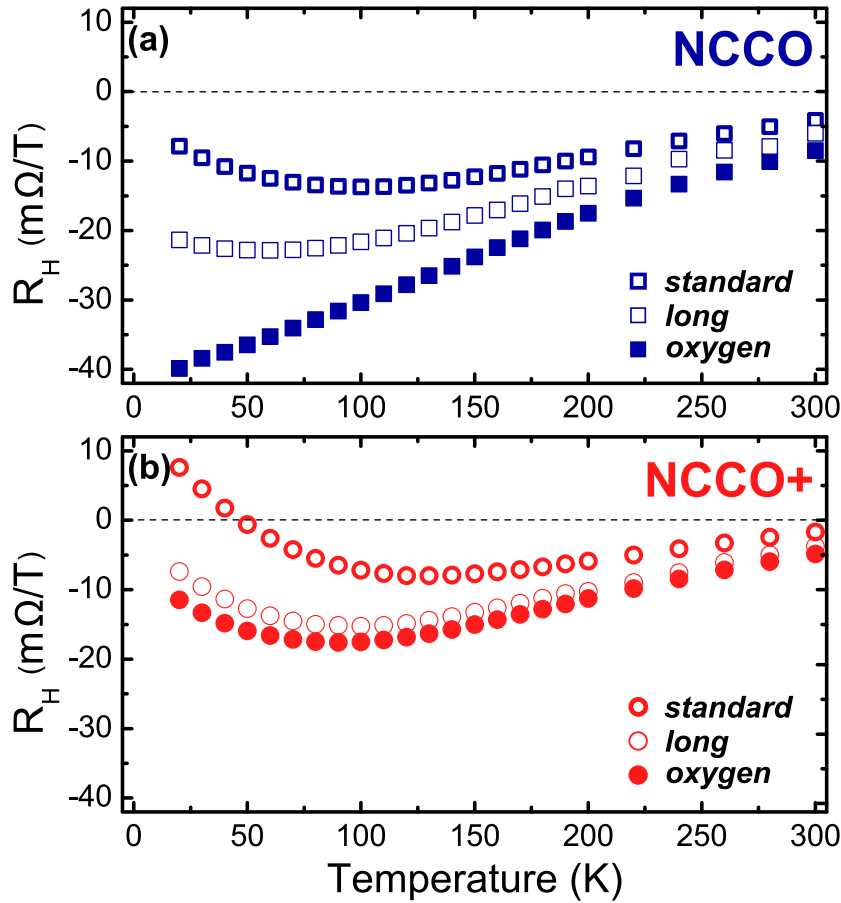}
\caption{Hall measurements on 70 nm \NCCOopt films deposited from the stoichiometric target (a) and the non-stoichiometric target with extra copper added (b) for the three different annealing procedures as described in table \ref{tab:ann}: {\em standard} (half filled symbols), {\em long} (open symbols) and {\em oxygen} (closed symbols).}
\label{fig:5}

\end{figure}

The Hall effect measurements confirm the observations from the sheet resistance measurements, see figure \ref{fig:5}(a,b). The \NCCOopt films grown with the standard target show a wide spread in the Hall coefficient, especially at low temperatures, where the characteristic minimum has completely disappeared for the films annealed following the {\em oxygen} procedure, see figure \ref{fig:5}(a). This was also reported for oxygenated films of \NCCO \cite{Jiang1994} and \PCCO \cite{Gauthier2007}. For the \NCCOopt films with extra copper we see a closer grouping of all the curves and they all show the characteristic minimum, see figure \ref{fig:5}(b).  It is interesting to note that both a longer vacuum annealing and an oxygen annealing following a short vacuum annealing suppresses the cross-over to a positive Hall coefficient, not depending on the presence of the parasitic \CNO phase. In the latter case this can be explained by the counter-doping effect of oxygen, effectively shifting the doping away from optimal doping, towards underdoping \cite{Higgins2006}. The former case is not immediately obvious. The cross-over to positive Hall coefficients for the extra copper films with an 8 minute vacuum annealing, suggests that the samples are already at optimal doping or even overdoped.  It has been shown that, during vacuum annealing, oxygen is removed from the CuO\subs{2} planes, suppressing anti-ferromagnetic ordering and promoting hole-like carriers \cite{Richard2004,Riou2004,Gauthier2007,Song2012}. This is accompanied by an increased normal-state resistance for over-reduction \cite{Jiang1994,Gupta1989,Brinkmann1996}, attributed to increased defect and impurity scattering \cite{Song2012}. Annealing for 45 minutes should therefore only shift the Hall coefficient towards more hole-like character, as is commonly observed for oxygen reduction in \NCCO \cite{Jiang1994,Xu1996,Mao1995}. This is not what we observe, we do see an increased normal-state resistance, but the Hall coefficients shifts toward more negative values. We suggest that as more and more oxygen is removed from the CuO\subs{2} planes, scattering is increased. As the Hall coefficient for a two-band system is also linked to the mobility of the carriers, our observations can be explained by the scattering centers having more influence on the mobility of the hole-like carriers than on the electron-like carriers. This would explain both the higher normal state resistivity and the shift to a more negative Hall coefficient.

\section{Conclusions}
Our experiments show that the addition of extra copper to the PLD target of \NCCOopt can increase the quality and $T_c$ of thin films by suppressing the formation of the parasitic \CNO phase. The presence or absence of the parasitic \CNO phase in \NCCOopt films has a strong influence on $T_c$ and the transport properties above $T_c$. For long reduction the difference is minimal, while the biggest difference is observed for strong oxygenation.   The relative minor influence of the exact annealing procedure on $T_c$ and the Hall coefficient for \NCCOopt films grown with extra copper in the target suggests that studies into the exact role of oxygen in the reduction process for {\em n-}type cuprates should not overlook the influence the presence of the \CRO parasitic phase can have. We have also shown that the \NCCOopt films with extra copper can even be subjected to a 1 bar oxygen annealing suitable for the growth of {\em p-}type cuprates such as \LSCO and \YBCO. This makes \NCCO grown from a non-stoichiometric target with extra copper a prime candidate for studying {\em pn}-physics in the cuprate superconductors, opening the way to integrate electron-doped cuprates with their hole-doped counterparts on a single chip. Such combinations are of interest for example for the creation of superconducting {\em pn}-junctions or to explore electron-hole interactions in the rich phase diagram of the cuprates.

\ack
This research was supported by the Dutch NWO foundation through a VICI grant, XRW is supported by a NWO Rubicon grant (2011, 680-50-1114). The authors thank E.G. Keim for TEM sample preparation and imaging, and P. Fournier, S. Harkema and F.J.G. Roesthuis for valuable discussion.

\end{document}